\documentstyle[epsf]{l-aa}

\def\ddx#1#2{{d #1 \over d #2}} 
 
\def\Etal#1{et al.\ (\cite{#1})}	
\def\etal#1{et al.\ \cite{#1}}
\def\ie{i.e.}
\def\fig#1{Fig.\,\protect\ref{#1}}
\def\Fig#1{Figure\,\protect\ref{#1}}
\def\tab#1{Table\,\protect\ref{#1}}
\def\eq#1{Eq.\,\protect(\ref{#1})}

\def\sec#1{Sect.\,\protect\ref{#1}}
\def\Sec#1{Section\,\protect\ref{#1}}

\def\App#1{Appendix\,\protect\ref{#1}}

\def\ncno{{\rm nCNO}}

\def\eg{{e.g.}}

\def\SRM{seismic reference model}
\def\OD{opacities}
\def\EOS{equation of state}
\def\SMD{solar models with microscopic diffusion}
\def\SM{solar models without microscopic diffusion}
\def\SMSMD{solar models with and without microscopic diffusion}
\def\PMS{pre-main sequence}
\def\MS{main sequence}
\def\ZAMS{zero-age main sequence}
\def\hZAMS{homogeneous \ZAMS}
\def\ML{mass loss}
\def\MLR{\ML\ rate}
\def\mMLR{mild \MLR}
\def\sMLR{strong \MLR}
\def\MI{$\cal Z$}

\begin{document}

\thesaurus{06.01.1, 06.05.1, 06.15.1, 06.09.1}

\title{Updated Solar Models}

\author{P. Morel\inst{1}, J. Provost\inst{1} and G. Berthomieu\inst{1}}

\offprints{P. Morel, morel@obs-nice.fr}

\institute{D\'epartement Cassini, UMR CNRS 6529, Observatoire de la C\^ote 
d'Azur, BP 4229, 06304 Nice CEDEX 4, France
}

\date{Received date; accepted date}

\maketitle

\begin{abstract}
Solar models computed with mass loss, microscopic diffusion of helium
and heavy element, and with updated physics
have been evolved from the pre-main sequence to present day;
they are compared to the observational constraints including lithium depletion
 and to
the \SRM\ of Basu \Etal{bcd}, derived by inversion.
Microscopic diffusion
significantly improves the agreement with the observed solar frequencies
and agree with the \SRM\ within $\pm 0.2\%$ for the sound velocity and 
$\pm 1\%$ for the density, but slightly worsens the neutrino problem. 
Neither microscopic diffusion nor
overshooting explain the observed lithium depletion consistently
with helioseismological constraints, while a mass loss process does it.
Models computed with OPAL \EOS\ and \OD\ are  in a better agreement
with the seismic sound speed.
To reach the level of precision of helioseismological observations 
the accuracy of solar models still needs to be improved by
one magnitude; any such improvement will necessitate equation of state and 
opacity data taking into account of detailed changes in the mixture. 
\end{abstract}


\section{Introduction}
The successful development of helioseismology imposes very strong
 constraints on the structure of the interior of solar models and 
has led solar modelers to improve the physical inputs of their models.
One important step is the insertion of microscopic diffusion.
The \SMD\ (Proffitt \& Michaud \cite{pm}; 
Bahcall \& Pinsonneault \cite{bpb}; Christensen-Dalsgaard \etal{cpt};
Kovetz \& Shaviv \cite{ks}; Proffitt \cite{pr}; 
Morel \etal{mli}; Gabriel \& Carlier \cite{gmc};
Degl'Innocenti
\etal{ddfr}; Brun \etal{blmt}) closely agree with constraints
of the solar interior inferred by
helioseismology as reviewed by Christensen-Dalsgaard \Etal{daa}.
The recent improvements of \EOS\ (Rogers \etal{rsi})
and \OD\ (Iglesias \& Rogers \cite{ir}) have brought the difference between
the solar sound speed of the seismic model of Basu et al.
(\cite{bcd}), from LowL data (hereafter, "\SRM") and
the sound speeds computed from present day solar models within a rms
discrepancy better than 0.2\% (Gough \etal{get});
the values of the radius at the bottom
of the convection zone (Christensen-Dalsgaard \etal{cgt}) and the helium 
abundance
at the solar surface (P\'erez Hern\'andez \& Christensen-Dalsgaard \cite{hc};
Antia \& Basu \cite{ab}; Basu \& Antia \cite{ba})
predicted by models agree within the error bar with their values
inferred from helioseismology (\eg, Basu \cite{b}).
Nevertheless the large excesses of
neutrino rates (\eg, Bahcall \cite{b}) and of lithium abundance predicted
by the models still appear in
strong conflict with the good agreement between the structure of the model and 
the inferred sound speed and density.
Up to now, all attempts to mix the Sun in such a way as to minimize the
discrepancies between solar observations and predictions for, either neutrinos
or lithium depletion, are ruled out by helioseismology.
So mixing may be not the relevant process; it
appears thus that for the neutrino problem the discrepancies are in pass to
be understood rather through a modest extension of the electroweak theory 
(Bahcall \& Krastev 1996)
than through improvements or change of physics.
Promising processes for accounting for the observed lithium depletion
are mass loss (Boothroyd \etal{bsf}; 
Guzik \& Cox \cite{gc}; Henyey \& Ulrich \cite{hu}) and turbulent
diffusion induced by rotation (Charbonnel \etal{cvz},
Chaboyer \etal{cdp}, Richard \etal{rvcd}).

In this work, we have reconsidered the effect of microscopic diffusion and
discussed the estimation of the heavy element content used in opacities calculations. Starting from 
chemically homogeneous \PMS\ models
we have computed calibrated \SMSMD. We have considered two different mass loss
laws, undershooting and overshooting of convection zones,
and also examined the effects of different \EOS\ and \OD. However
turbulent diffusion induced either by
rotation 
or internal waves (Schatzman \cite {s}, 
Montalban \& Schatzman \cite{ms}) is ignored.

The models have been compared to the
\SRM\ through the quantities $(c_\odot-c_{\rm model})/c_\odot$ and 
$(\rho_\odot-\rho_{\rm model})/\rho_\odot$,
$c_{\rm model}$, $c_\odot$, $\rho_{\rm model}$ and $\rho_\odot$ 
are, respectively, the sound speeds and the densities of the
model and of the \SRM. The oscillation frequencies of the
solar models have been compared to the GONG data observations. 
Likewise, the neutrino fluxes expected for the
three experiments and the lithium depletion at present age have
been computed and compared with the measurements.
These solar models are updated and improved versions of
the models discussed in Berthomieu \Etal{be}.
 
In \Sec{sec:gc} global parameters and helioseismological constraints
are briefly recalled for references and notations. In
\Sec{sec:phy} the physics is described. Outlines of numerical
techniques used so far are given in \Sec{sec:com}. Results and discussion
are presented in
\Sec{sec:r} and the conclusions in \Sec{sec:dis}. Two appendices are devoted to
detailed discussions of heavy element abundances which enter in \EOS\ and \OD.

\section{Observational constraints and calibration of solar models}
\label{sec:gc}
For reference, the solar global parameters, at present,
and the constraints inferred
from helioseismology are given in \tab{tab:1} were the standard notations and
the units are recalled in the caption. Recent observed neutrino fluxes at earth 
from the Gallium $\Phi_{\rm Ga}$,
Chlorine $\Phi_{\rm Cl}$ and Kamiokande $\Phi_{\rm Ka}$ experiments
are reported. The frequency differences $\overline{\delta\nu}_{02}$
and $\overline{\delta\nu}_{13}$ which correspond
to  mean values of recent observations from the ground based networks
IRIS and BiSON and from the spatial experiments VIRGO and GOLF are given (see 
section 4.2).
There $Y_\odot$ is the surface abundance per unit mass of helium,
 determined  by Basu \& Antia (1995) with respectively MHD and OPAL
 equation of state, and $R_{\rm ZC}$ is the radius at the bottom of
 the convection zone.
The solar age given in Table 1 corresponds to the start of
 main sequence  evolution.  We have adopted   
an  evolutionary time of  4.5\,Gyr for models starting from the \ZAMS\ and
  4.55\,Gyr for models including \PMS.
\begin{table}
\caption[ ]{Solar data at present day; 
the units are, $10^{33}$ g for the mass $M_\odot$, $10^{10}$ cm for
the radius $R_\odot$, $10^{33}$ erg\,s$^{-1}$ for the luminosity $L_\odot$,
Gyr for the age $t_\odot$,
dex ($^1$H=12) for the surface depletion of lithium Li$_\odot$, 
SNU for $\Phi_{\rm Ga}$ and $\Phi_{\rm Cl}$, ev day$^{-1}$ for $\Phi_{\rm Ka}$,
$\mu$Hz for the frequency differences $\overline{\delta\nu}_{02}$ and 
$\overline{\delta\nu}_{13}$.
}\label{tab:1}
\begin{tabular}{lll}
 \hline
$M_\odot$       & $1.9891\pm 0.0004$ & Cohen \& Taylor (\cite{ct}) \\
$R_\odot$       & $6.9599\pm 0.001$  & Guenther \Etal{gdkp} \\
$L_\odot$       & $3.846\pm 0.005$   & Guenther \Etal{gdkp} \\
$t_\odot$       & $4.52\pm 0.4$     & Guenther \Etal{gdkp} \\
$(Z/X)_\odot$   & $0.0245\times(1\pm 0.1) $ & Grevesse \& Noels (\cite{gn})\\
$Y_\odot$       & 0.246 ; 0.249 & Basu \& Antia (1995)  \\
$R_{\rm ZC}/R_\odot$ & $0.713\pm 0.003$ & Christensen-Dalsgaard \\
&&\Etal{cgt}\\
Li$_\odot$      & $1.16\pm 0.1 $ & Anders \& Grevesse (\cite{ag})\\
$\Phi_{\rm Ga}$ & $69 ^{+7.8}_{-8.1} $ & Hampel \Etal{ha}\\  
$\Phi_{\rm Cl}$ &$2.55\pm 0.23 $ & Davis (\cite{da})\\
$\Phi_{\rm Ka}$ & $0.29\pm 0.02$ &Fukuda \Etal{fu}\\
$\overline{\delta\nu}_{02}$& $9.01\pm 0.05$ & Provost (\cite{p})\\
$\overline{\delta\nu}_{13}$& $15.90\pm 0.08 $ & Provost (\cite{p})\\ \hline 
\end{tabular}
\end{table}
The amount of heavy element per unit 
of mass for present day $Z_\odot=0.0181(1\pm 0.1)$ is derived from $Y_\odot$
and from $(Z/X)_\odot$ the ratio of heavy element to hydrogen measured at the
surface.

The solar models discussed in this paper are calibrated within
a relative accuracy better than $10^{-4}$ by adjusting:
i) the ratio $l/H_{\rm p}$ of the mixing-length to the 
pressure scale height, ii) the protosolar mass fraction $X_{\rm p}$
of hydrogen, iii) the protosolar mass fraction $(Z/X)_{\rm p}$ of
heavy element to hydrogen and, iv)
the mass $M_{\rm p}$ of the primitive Sun,
in order that, at present day, the models have the values of
\tab{tab:1} for, i) the luminosity, ii)
the radius, iii) the mass fraction of heavy element to hydrogen
and, iv) the solar mass.

\section{Physics of Solar models}\label{sec:phy}
\subsection{Initial conditions and nuclear network}\label{sec:pms}
Here the evolution of solar models, but one, include
the \PMS; they are initialized with chemically homogeneous models,
according to the procedure of Iben (\cite{ib}) using a {\em fixed} contraction 
factor
$c\equiv 0.02L_\odot M_\odot^{-1}K$, see details in Morel (\cite{mp96}).

The general nuclear network we used
contains the following  twelve species 
$^1$H, $^2$H, $^3$He, $^4$He, $^7$Li, $^7$Be, $^{12}$C, $^{13}$C, 
$^{14}$N, $^{15}$N, $^{16}$O and $^{17}$O which enter into
the most important nuclear reactions of the PP+CNO cycles.
The relevant nuclear reaction rates
are taken from the tabulations of Caughlan and Fowler (\cite{cf}). Weak 
screening is assumed\footnote{It has been recently demonstrated that weak 
screening is a very good approximate of the exact solution
of the Schr\"odinger equation for the fundamental
PP reaction (Bahcall \etal{bck}).}. The abundance of any chemical is explicitly 
computed \ie, no element is assumed to be at equilibrium.
The protosolar abundances of deuterium
\footnote{These computations do not 
use the re-estimated value of the protosolar $^2$H/$^1$H ratio by
Gautier \& Morel (\cite{gm}).}
and of species heavier than helium are calculated using
their nuclide ratios with respect to hydrogen 
(Anders \& Grevesse \cite{ag});
the protosolar mass fraction $Y_{\rm p}$ of helium is taken as: 
$Y_{\rm p}\equiv 1 - X_{\rm p}(1+(Z/X)_{\rm p})$; 
the protosolar $^3$He abundance is derived 
from the protosolar $^4{\rm He}\equiv Y_{\rm p}$ abundance according to the 
isotopic ratio
$^3{\rm He}/^4{\rm He}=1.42\times10^{-4}$.
 For lithium, we have used the
ratio by number $(^7{\rm Li}/^1{\rm H})_{\rm p}=3.277$ dex;
the unknown protosolar ratio by number $^7$Be/$^1$H is
taken to $-3.58$ dex \ie, about its equilibrium value at center of the Sun.
As a definition, we have considered that a \PMS\ model becomes a
\MS\ model as soon as 99\% of the energy released is from nuclear;
this transition occurs
at an age between 40\,Myr and 80\,Myr.
At that time, due to the CNO nuclear energy produced by
the conversion of protosolar $^{12}$C onto $^{14}$N, the models have a 
convective core of typical extent $\sim 10\%$ of the total radius and
the outer solar convection zone has already receded about to its present
day location. We emphasize the importance of \PMS\ evolution for the study 
of the depletion of light element; with physical conditions of solar \hZAMS\ 
and with the nuclide abundances of $^2$H and $^7$Li,
the two reactions $^2$H($p,\gamma)\,^3$He and $^7$Li($p,\alpha)\,^4$He
are strongly far from equilibrium;
a \hZAMS\ model can be found only if the abundances of $^2$H and $^7$Li are
taken near to their equilibrium values,
that is inappropriate for studies of light element depletion.

\subsection{The \OD\ and \EOS}\label{sec:eos}
We have used the \OD\ of Livermore Library (Rogers \& Iglesias
\cite{ri}; Rogers \etal{rsi}) extended with the low temperature opacities
of Kurucz's (\cite{ku}) interpolated with a birational spline interpolation
(Houdek \& Rogl \cite{hr}) and, either the CEFF \EOS\ 
(Christensen-Dalsgaard \& D\"appen \cite{cd}),
or the OPAL \EOS\ (Rogers \etal{rsi}). Because it depends not much on $Z$
and to avoid too heavy numerical computations,
we have fixed the amount of heavy element in the \EOS\ to values closed to
their mean amounts in calibrated models namely, $Z\equiv 0.0175$
for \SM\ and $Z\equiv 0.0190$ for \SMD.
The \OD\ and \EOS\ are tabulated for mixtures with various amounts
of hydrogen, $X_{^1{\rm H}}$, but with {\em fixed}
ratios between the species of heavy element. During the evolution 
these ratios are modified by the 
thermonuclear reactions and by the microscopic diffusion and the new mass 
fraction
of each isotope needs to be re-estimated as detailed in \App{ap:1}. 
In \MS\ solar models a local maximum of $^3$He occurs around $R\sim 
0.3R_\odot$, it amounts to 10\% of the local helium content
$X_{^3{\rm He}}\sim 0.1\times X_{^4{\rm He}}\sim Z$, here $X_{^3{\rm He}}$
and $X_{^4{\rm He}}$
are the mass fractions of, respectively, $^3{\rm He}$ and $^4{\rm He}$;
as seen in \App{ap:2} this effect needs to be
taken into account in the calculation of the amount of heavy element
used for the determination of \OD\ and \EOS.

\subsection{Diffusion}\label{sec:z}
The turbulent diffusion is ignored in the solar models discussed in this paper.
Various treatment of the microscopic microscopic diffusion
leads to diffusion rates in good agreement (Bahcall \& Pinsonneault \cite{bpc}).
We have used the microscopic diffusion coefficients of
Michaud \& Proffitt (\cite{mipr}).
All chemicals including either $X_\ncno$ or
$\cal X_{\rm Z}$ (see below), but $^1$H and $^4$He, are trace elements.
Equation (14) of Michaud \& Proffitt valid 
for a H-He mixture, is approximatively extended to a mixture including 
heavier elements, by writing (Alecian \cite{al}): 
\begin{equation}\label{eq:vd}
X_{^1\rm H}V_{^1\rm H}=-(1-X_{^1\rm H}-Z)V_{^4\rm He}
\end{equation}
instead of $X_{^1\rm H}V_{^1\rm H}=-(1-X_{^1\rm H})V_{^4\rm He}$,
therefore $X_{^1\rm H}+X_{^4\rm He}+Z\equiv 1$ even with  
microscopic diffusion,
here $V_{^1\rm H}$ and $V_{^4\rm He}$ are respectively
the diffusion velocities of hydrogen and helium.

The \OD\ and \EOS\ are functions of the heavy element content $Z$,
through the number of free 
electrons and the abundances of efficient absorbers which do not necessarily
belong to the nuclear network \eg, $^{56}$Fe.
Due to diffusion and nuclear reactions, $Z$ changes
as the Sun evolves, as well as the ratios between the abundances of chemicals;
in the following we shall designate by $Z_\kappa$
the amount of heavy element used
for the computation of \OD\ in models and we discuss next
three approximate estimates employed here.

For the calculation of \SM\ either, i) $Z_\kappa$ is given by \eq{eq:z} 
resulting of
detailed calculations of abundances including the changes
due to the nuclear reactions -- as detailed in \App{ap:1} -- 
but, the ratios between the
abundances are not consistent with the values used for the creations
of \EOS\ and \OD\ data or, ii) $Z_\kappa$ is kept to its protosolar
value $Z_\kappa=Cte=Z_{\rm p}$ \ie, one neglects the changes of abundances of 
heavy element.

For \SMD\ the alternative is more subtle,
in the core it is expected that the abundances
of the heaviest chemicals are magnified by gravitational settling with the
consequence of an enhancement of the opacity there (Turck-Chi\`eze 
\etal{tdfpsv}), therefore a careful
estimate of $Z_\kappa$ is required to infer the effects of this process.

A first possibility is to consider that, since opacity tables are available
only for one mixture of heavy element, we do not know how to take into account
properly their changes due both to nuclear reactions and diffusion, so we
just ignore them and take $Z_\kappa=Cte=Z_{\rm p}$.

A second possibility is to model the 
diffusion of heavy element as a whole, by means of a mean fictitious 
chemical ${\cal X}_{\rm Z}$, with the mean atomic 
weight $M_{\rm Z}$, and the mean charge $z_{\rm Z}$, of the heavy species;
${\cal X}_{\rm Z}$ being diffused as an extra
trace element but it is not affected by the thermonuclear reactions; then 
the ratios between the heavy element entering onto ${\cal X}_{\rm Z}$
are fixed, their changes due to thermonuclear reactions are ignored and
${\cal X}_{\rm Z}$ can differ from its value given by \eq{eq:z}.

We have also investigated a third possibility -- labeled \MI\ --
namely, $Z_\kappa$ is separated in two parts, the first consists of
the chemicals heavier than helium which, belonging to the CNO nuclear network,
are both diffused and nuclearly processed,
the second part consists of a fictitious mean non-CNO species $X_\ncno$,
of atomic weight $M_\ncno$ and charge $z_\ncno$,
which is only diffused (\eq{eq:z}). Hence in the estimate of $Z_\kappa$, enter into account
the changes of CNO abundances
caused by diffusion, nuclear processes and the
effects of the large gravitational settling of heaviest non-CNO
species but, the ratios 
between the chemicals can differ from the values
used for the computations of \OD\ and \EOS\ data. This third possibility is
similar to the treatment of Proffitt (\cite{pr}), where in addition,
an attempt is made -- namely its Eq. (3)  -- to approximate 
the changes of ratios between the heavy element in the calculation
of $Z_\kappa$.  

For the mean chemicals, ${\cal X}_{\rm Z}$ and
$X_\ncno$, we have only taken into account the heavy element with abundances 
larger than 
7.0 in the table 2 of Anders \& Grevesse (1989), namely: C, N, 
O, Ne, Mg, Si, Fe; we have found, for ${\cal X}_{\rm Z}$: $M_{\rm Z}=17$, 
$z_{\rm Z}=8$ and, for $X_\ncno$: $M_\ncno=27$ and $z_\ncno=13$.

\subsection{Convection, overshooting and undershooting}\label{sec:cv} 
The basic formulation of the standard 
mixing length theory is used.
The convection zones are delimited 
according to the Schwarzschild's criterion.
In convection zones and in their extensions by overshooting or undershooting, 
the chemicals are {\em simultaneously} homogenized and integrated with 
respect to time (Morel \& Schatzman \cite{ms}; 
Morel \cite{mp96}). The mixing includes every convective
zone and its extent by undershooting or overshooting where the
temperature gradient is kept equal to its adiabatic value (Zahn \cite{z}).
The amount of overshooting
of a convective core is taken as $\zeta_{\rm ov}\min(H_{\rm p}, R_{\rm cv})$,
$\zeta_{\rm ov}$ is the overshooting factor and
$R_{\rm cv}$ is the radius of the convective core; at the outer limit \ie,
in the atmosphere, no overshooting is allowed.
With an overshooting factor of
$\zeta_{\rm ov}\sim 0.3$, the young Sun has a convective core, 
up to the age $t\sim 1\,$Gyr; this results an
enrichment in nuclear fuel which does not lead to significant changes
in the present day solar model (Morel \etal{mli}).

\subsection{Lithium depletion and mass loss}\label{sec:li}
The depletion of 
lithium is mainly caused by the reaction $^7$Li($p,\alpha)\,^4$He.
For solar conditions ($T \la 1.5\times10^7$K) it is, at least, 10 times more
efficient than all the reactions involving $^7$Li,
except $^7$Li($d,n)2\,^4$He which 
is only two times less efficient (Caughlan \& Fowler \cite{cf}),
but the abundance of deuterium being almost zero,
this reaction depletes no lithium.
Therefore we can estimate that the abundance of $^7$Li resulting from our
calculations is $\sim 2\%$ over-estimated -- $^6$Li
destroyed around $2\times 10^6$K is not observed in the solar
photosphere (B\"ohm-Vitense \cite{bo}). 
\begin{figure*}
\epsfxsize =19.cm
\epsfbox[25 18 587 274]{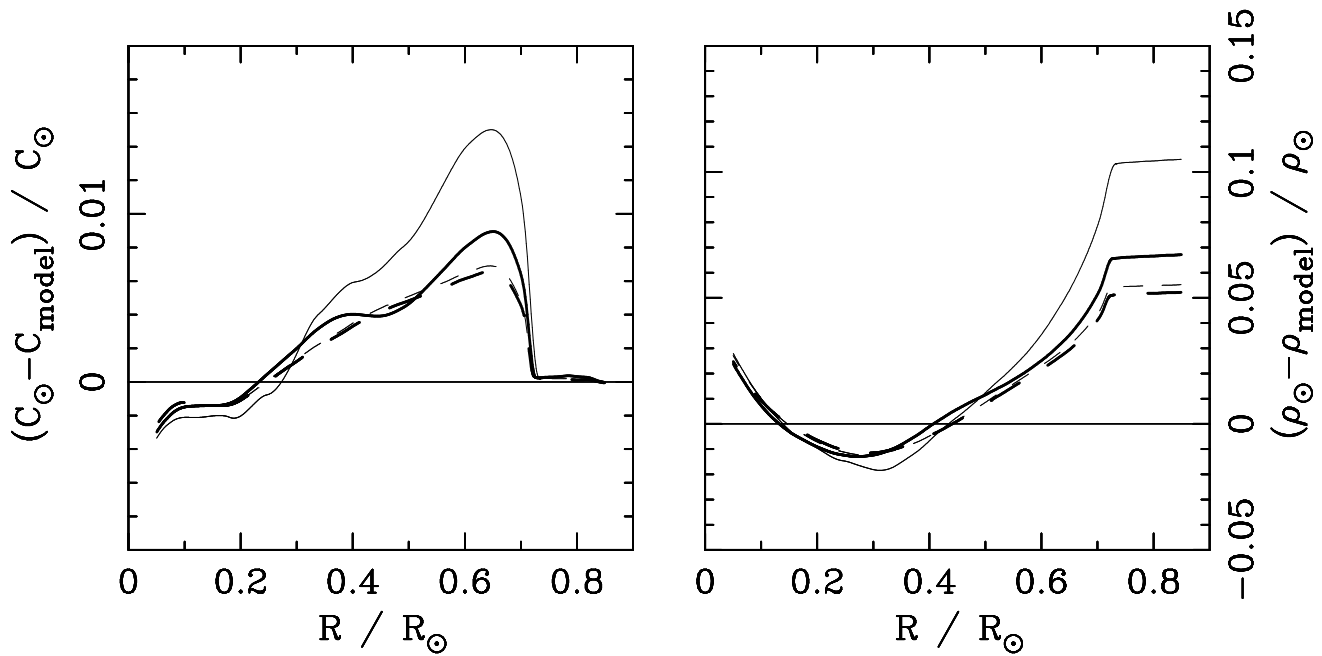}
\caption[ ]{Fractional differences for sound velocities (left) and
densities (right) between the \SM\ of
\tab{tab:11} and the \SRM\ of Basu \Etal{bcd}.
The main features of the physics used in each model are
schematically recalled with the notations of \tab{tab:31}.
S1 \{CEFF, OPAL92, PMS\} ({\sl thin full}),
S2 \{CEFF, OPAL95, PMS\} ({\sl thick full}),
S3 \{OPAL, OPAL95, PMS\} ({\sl thin dashed}),
S4 \{OPAL, OPAL95, ZAMS\} ({\sl thick dashed}).  
}\label{fig:ba1}
\end{figure*}

As suggested (Soderblom \etal{spfj}) by the observations of the Pleiades
(80-100\,Myr) for stars with masses closer to the solar value
only small lithium depletion seems to occur during the \PMS\ while,
from the observations of the Hyades (600\,Myr)
it appears that the most important amount of the 
depletion\footnote{despite the
higher metallicity of Hyades.} should occur during the
first $\lse 500$\,Myr of the \MS.
Such a depletion can result from mass loss: in a star of mass
$M\sim 1.1M_\odot$,
at the onset of \MS\ the convection zone recedes at
mass $m>1M_\odot$, the layers which correspond
to the location of the present day convection zone \ie,
$m\in [\sim 0.97M_\odot, \sim 1 M_\odot]$, are radiative and the temperature
there is large enough to deplete the lithium.
A mass loss of $\sim 0.1M_\odot$,
occurring at that time ejects the outer layers, and the present day
convection zone is formed by mixing these
lithium poor shells, with the result of
the observed enhancement of lithium depletion. 
Following Guzik \& Cox (\cite{gc}) we have studied two laws of
\MLR\ namely, a \mMLR:
\begin{equation}\label{eq:mmlr}
\dot M=-2\times10^{-10}\exp(-\frac t{0.45})\ M_\odot\ {\rm yr}^{-1},
\end{equation}
and a \sMLR:
\begin{equation}\label{eq:smlr}
\left\{
\begin{array}{ccl}
\dot M&=&-5\times10^{-10}\ M_\odot\ {\rm yr}^{-1}\ {\rm if}\ M(t)>M_\odot, \\
\dot M&=&0\ {\rm otherwise;}
\end{array}
\right.
\end{equation}
here $t$, in Gyr, is the age of the model and $M(t)$ is its mass at time $t$;
with the \mMLR\ the 
protosolar mass $M_{\rm p}\sim 1.1M_\odot$ is adjusted to fit the
present day solar mass.
 
\section{Computation of models and frequencies}\label{sec:com}
\subsection{Solar models}
The solar models discussed in this paper have been computed
using the code CESAM (Morel \cite{mp96}); along the evolution atmospheres
are restored (Morel \etal{mp94}). 
The number of mass shells and the 
time steps are fixed according to numerical criteria; at the onset
of calculations (\PMS) the number of integration points is typically
of the order of 550
and around 1030 for the present day solar models ($\sim 950$ for the core
and the envelope, $\sim 80$ for the atmosphere); 
the whole evolution involved about 140 models, with 60 models from \ZAMS\ to 
present day.
It has been checked that the numerical internal accuracy on sound speed
is better than $5\times10^{-4}$ for the whole model, but differences below
that level are perhaps meaningless due to uncertainties of interpolated physical
data. It is assumed that the lost mass is simply detached from the 
star by some process which is not described.
We also emphasize on the fact that in our calculations, the $^7$Li
depletion results from
{\em simultaneous} computations of chemical changes
due to nuclear reactions, diffusion and convective mixing.
The estimated fluxes for the three neutrino
experiments, $\Phi_{\rm Cl}$, $\Phi_{\rm Ga}$ and $\Phi_{\rm Ka}$
are computed according to Berthomieu \Etal{be}.
 
\subsection{p-Mode Oscillation Calculations}
The frequencies of linear, adiabatic, global acoustic modes of the
solar models have been computed for degrees $\ell$=0 to 150
and have been compared to the observations.
For the computation of frequencies the models 
are extended to about 1800 shells.
The characteristic low degree p mode frequency
differences $\Delta\nu_{n,\ell}=\nu_{n,\ell}-\nu_{n-1, \ell+2}$ 
for $\ell$=0 and 1,
which provide information on the properties of the core of the Sun,
have been fitted by linear regressions with respect to $n$: 
\[
\Delta\nu_{n,\ell}= \delta\nu_{n,\ell} + S_\ell (n-n_0), 
\]
with  $n_0=21$ and for $\ell =0,1$, 
both for the observations and the theoretical frequencies.
The quantities $\overline {\delta\nu}_{02}$ and 
 $\overline {\delta\nu}_{13}$ given in Table 1 are weighted mean values 
of the $\delta\nu_{n,\ell}$ derived from BiSON (Chaplin \etal{ch}),
IRIS (Gelly \etal{ge}), GOLF (Grec \etal{golf}) and VIRGO (Fr\"ohlich 
\etal{vir}). Despite the fact that solar activity is not expected to have
a large influence on the solar core structure, we have also represented
in \fig{fig:dnu} another mean restricted to low activity observations.
 
For the gravity modes which have not yet been observed, we give the 
characteristic
asymptotic spacing period $P_0$ according to Provost \& Berthomieu (\cite{pb}).

\section{Results and discussion}\label{sec:r}
We have computed \SM, (labeled "S$n$", $n=1,\ldots 4$) and
\SMD\ (labeled "D$n$", $n=1,\ldots, 12$)
using the physics and physical data described \sec{sec:phy} and summarized
in \tab{tab:31} for each model.

\begin{table*}
\caption[]{
Summary of physical inputs for \SMSMD\ of \tab{tab:11} and \tab{tab:21}.
For each model the
flag signals the physical option employed.
In the first three lines, the evolutionary time is mentioned; the two next,
indicate either, if the evolution includes the \PMS\ and then the nuclear
network with 12 species is used or, if the evolution is initialized from the
\hZAMS\ then it uses the simplified nuclear network with $^2$H, $^7$Li and $^7$Be
at equilibrium. In the four next lines the flags indicate if overshoot
(Ov-shoot),
undershoot (Un-shoot), \mMLR\ (mMLR) or \sMLR\ (sMLR) are allowed for. In the next three lines
is indicated how the amount of heavy element is computed for the calculation of
\OD\ (see text). In the last lines the mass ratio of heavy element to hydrogen,
used for the calibration, the \OD\ and the \EOS\ used are given.  
}\label{tab:31}
\begin{tabular}{lccccccccccccccccc} \\  \hline \\
                        &S1&S2   &S3&S4 &D1&D2&D3 &D4&D5&D6&D7&D8&D9 &D10&D11&D12& \\ \\ \hline \\
age 4.65\,Gy            &  &     &  &   &  &  &   &  &  &  &  & *&  *&   &   &   &\\                    
age 4.55\,Gy            & *& *   & *&   &* &* & * & *& *& *& *&  &   & * & * & * &\\
age 4.50\,Gy            &  &     &  &*  &  &  &   &  &  &  &  &  &   &   &   &   &\\
PMS 12 Chem.            & *& *   & *&   & *& *& * & *& *& *& *& *&  *& * & * & * &\\
ZAMS 9 Chem.            &  &     &  & * &  &  &   &  &  &  &  &  &   &   &   &   &\\
Ov-shoot                &  &     &  &   &  &  &   &  &  &  & *&  &   &   &   &   &\\
Un-shoot                &  &     &  &   &  &  &   &  &  & *&  &  &   &   &   &   &\\
mMLR                    &  &     &  &   &  &  &   &  & *&  &  &  &   &   &   &   &\\
sMLR                    &  &     &  &   &  &  &   & *&  &  &  &  &   &   & * &   &\\
$Z_\kappa=Z_{\rm p}=Cte$& *& *   &* & * & *&  &   &* &* & *&* &* & * & * &   &   &\\
$Z_\kappa$=\MI          &  &     &  &   &  &  &   &  &  &  &  &  &   &   & * & * &\\
$Z_\kappa={\cal X}_{\rm Z}$   &  &     &  &   &  & *& * &  &  &  &  &  &   &   &   &   &\\
$(Z/X)_\odot$=0.0245    & *&  *  & *&  *& *& *& * & *& *& *& *& *&   &  *& * & * &\\
$(Z/X)_\odot$=0.0260    &  &     &  &   &  &  &   &  &  &  &  &  &  *&   &   &   &\\ 
opa-OPAL92              & *&     &  &   &  &  &   &  &  &  &  &  &   &   &   &   &\\
opa-OPAL95	        &  &  *  & *& * & *&* & * & *& *& *& *& *&  *& * & * & * &\\
EOS-CEFF                & *&  *  &  &   & *& *&   & *& *& *& *& *& * &   &   &   &\\
EOS-OPAL                &  &     & *&  *&  &  & * &  &  &  &  &  &   & * & * & * &\\
\\ \hline \\
\end{tabular}
\end{table*}

\begin{table}
\caption[]{
Global characteristics of \SM.
$l$ is the mixing-length, $H_{\rm P}$ the pressure scale height;
$Y_{\rm p}$, and $X_{\rm nCNO}$ are the protosolar abundance in mass of
respectively, helium and non-CNO heavy element; $(Z/X)_{\rm p}$ is
the protosolar ratio, in mass, of heavy element to hydrogen;
Li$_{\rm ZAMS}$ and Li$_\odot$ respectively are the surface depletions
in dex (H$\equiv 12$) of $^7$Li at \ZAMS\ and at present day;
$Y_\odot$, $Z_\odot$ and $R_{\rm ZC}$ respectively are, at present day,
the surface abundances, per unit of mass, of helium and of heavy
element and the radius, in solar units at the bottom of
the convection zone;
$T_{\rm c}$, $\rho_{\rm c}$, $Y_{\rm c}$ and $Z_{\rm c}$ are the central
values at present day respectively of, the temperature
in units of $10^7$K, the density in g\,cm$^{-3}$ the abundances,
per unit of mass of helium and of heavy element. 
$\Phi_{\rm Ga}$ and $\Phi_{\rm Cl}$ in SNU and
$\Phi_{\rm Ka}$ in events day$^{-1}$, are the expected fluxes
for the three neutrino
experiments namely Gallium, Chlorine and Kamiokande;
$\delta\nu_{02}$ and $\delta\nu_{13}$ are the average values
in $\mu$Hz of the frequency
differences between the radial p-modes of degree $\ell=0-2$ and $\ell=1-3$.
$P_0$ is the characteristic spacing period of g modes in minutes.
Data of \tab{tab:1} for the Sun ($\odot$) are also recalled.
The physics used in the models is summarized in \tab{tab:31}. 
}\label{tab:11}
\begin{tabular}{lllllllllllllllllll} \\  \hline \\
            & $\odot$ &  S1  &  S2  &  S3  & S4   &\\ \\ \hline \\
$l/H_{\rm p}$   &     &1.78  &1.80  &1.78  & 1.78 &\\
$Y_{\rm p}\cong Y_\odot$& 0.246 &0.269 &0.271 &0.268 & 0.268&\\
$Z_{\rm p}\cong Z_\odot$&     &0.0175&0.0174&0.0175&0.0175&\\
$(Z/X)_{\rm p}$ &     &0.0245&0.0245&0.0245&0.0245&\\
$X_{\rm nCNO}$  &     &0.0035&0.0035&0.0035&0.0035&\\ \\
Li$_{\rm ZAMS}$ &     & 3.18 & 3.15 & 3.08 &      &\\ \\
Li$_\odot$      &1.12 & 3.18 & 3.15 & 3.08 &      &\\ \\
$R_{\rm ZC}$     &0.713&0.733 &0.726 &0.723 &0.723 &\\ \\
$T_{\rm c}$     &     &1.535 &1.539 &1.538 & 1.537&\\
$\rho_{\rm c}$  &     &146.3 &147.1 &146.5 &146.3 &\\
$Y_{\rm c}$     &     &0.615 &0.619 &0.617 &0.616 &\\
$Z_{\rm c}$     &     &0.0180&0.0179&0.0180&0.0180&\\ \\
$\Phi_{\rm Ga}$ & 79  & 120  & 121  & 121  & 121  &\\ 
$\Phi_{\rm Cl}$ &2.55 & 6.11 & 6.30 & 6.22 & 6.20 &\\ 
$\Phi_{\rm Ka}$ & 0.28& 0.47 & 0.48 & 0.48 & 0.48 &\\ \\
$\delta\nu_{02}$& 8.99& 9.35 & 9.30 & 9.34 & 9.30 &\\
$\delta\nu_{13}$&15.87& 16.42& 16.37& 16.40& 16.36&\\ 
\\
$P_{0}$         &     & 36.51& 36.48& 36.45& 36.51&\\
\\ \hline \\
\end{tabular}
\end{table}

\subsection{The \SM}
The global characteristics of \SM\ are summarized in \tab{tab:11}.
With respect to
our previous \SM\ (Berthomieu \etal{be}) illustrated by S1, the changes
are mainly due to \EOS\ and \OD.
\Fig{fig:ba1} plots the comparisons with the \SRM; there
the fractional differences for the sound speeds and densities
are plotted with respect to the radius. 
Globally the models S are hardly closer than 
$\sim\pm 1\%$ for the sound speed and $\sim\pm 10\%$ for the density to the 
\SRM.
The comparisons between \{S1, S2\} and the \SRM, show that the
improvement of models towards the \SRM\ is mainly due to the new \OD.

\begin{figure}
\epsfxsize =8.cm
\epsfbox{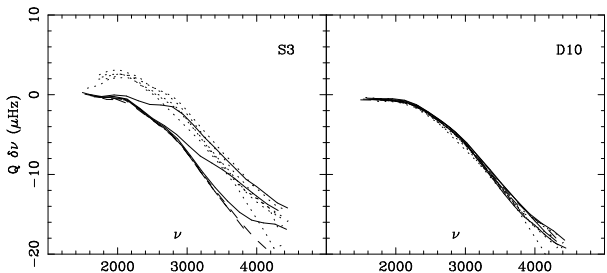}
\caption[ ]{
For models S3 and D10, the normalized frequency differences
between the GONG frequencies and the theoretical frequencies
are given for different degrees as a function of the frequency.
For each degree, the points are linked by a continuous line:
$\ell$~=~2, 3, 4, 5, 10, 20 ({\sl dotted});
$\ell$~=~30, 40, 50 ({\sl full}) and
$\ell$~=~70, 100, 120, 140 ({\sl dashed}).
}\label{fig:go}
\end{figure}

Both frequency differences $\delta\nu_{02}$ and
$\delta\nu_{13}$ are far from the observed values by more than 3$\sigma$
-- see \tab{tab:11}. The comparison with the GONG
frequencies and the theoretical frequencies for S3,  plotted in 
\fig{fig:go}, shows a 
noticeable dispersion of the curves for low and large degree modes,  due to
the bump ($\sim +0.6\%$) observed in sound speed below the convective zone on
\fig{fig:ba1}.

\paragraph{Consequence of the \PMS.}
In order to investigate the effects of the \PMS\ quasi-static contraction,
the model S4 has been computed with the species $^2$H, $^7$Li and $^7$Be at
equilibrium, initialized as a \hZAMS\ model and evolved 4.5\,Gyr \ie, 
$\simeq 50$\,Myr less than the models including the \PMS.
As reported in \tab{tab:11}
the global characteristics of S3 and S4 are similar and only a slight
difference on 
$\delta\nu_{02}$ and $\delta\nu_{13}$ reveals a very little change on
the seismic properties of the models.
The sound velocity of the two models are almost identical (see Fig.1), 
leading to normalized differences of frequencies between the two models 
 of the order of 0.5 $\mu$Hz.
Similar comparisons\footnote{not detailed in \tab{tab:31} for sake of 
briefness.}
performed for \SMD\ give differences of about 1 $\mu$Hz. Thus we can conclude
that
\PMS\ evolution and general nuclear network are a necessity
only for the study of specific relevant processes \eg, the lithium depletion.

\subsection{Standard \SMD}
The global characteristics of \SMD\ are summarized in \tab{tab:21}.
The comparisons of models with the \SRM\ are illustrated 
by \fig{fig:ba2} to \fig{fig:ba4} where
the fractional differences for the sound speeds and densities
are plotted with respect to the radius for the interval
$[0.05R_\odot, 0.85R_\odot]$ (Basu \etal{bcd}).

The photospheric helium content of our models with OPAL equation of state is
of the order of 0.245,  smaller than the two determinations of Basu \& 
Antia (1995). For models with the CEFF equation of state this helium content
is  slightly increased.
For all models D, except model D6 with penetration of convective elements
in the inner radiative zone, the radius at the bottom of
the convection zone is within the error bars of the value inferred by
helioseismology (see \tab{tab:1}) in agreement with previous works.


At the center the microscopic diffusion increases the density and
reduces the available nuclear fuel, namely $^1$H; a consequence of the
calibration is a larger central temperature
for \SMD\ than for \SM; therefore, as noticed by numerous authors,
the predicted neutrino rates
which are mainly sensitive to nuclear parameters
(\eg, Turck-Chi\`eze \& Lopez \cite{tl}; Dzitko \etal{dtdl}) are slightly larger 
for
\SMD\ than for \SM\ by
$\sim 7\%$, $\sim 25\%$ and $\sim 25\%$ respectively for $\Phi_{\rm Ga}$,
$\Phi_{\rm Cl}$ and $\Phi_{\rm Ka}$.


\begin{figure}
\epsfxsize =9.cm
\epsfbox[21 19 579 365]{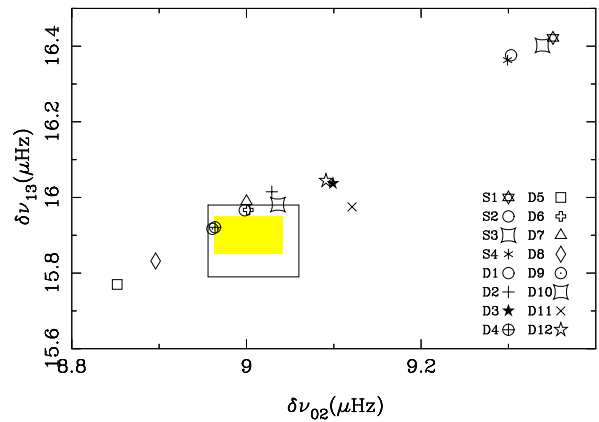}
\caption[ ]{Comparison of observed and theoretical estimations of the two mean
 frequency differences $\delta\nu_{02}$ and $\delta\nu_{13}$.
}\label{fig:dnu}
\end{figure}

In order to compare the seismic properties of the core of our models
 to those of
the Sun, we have plotted on \fig{fig:dnu} the values of the two mean
 frequency differences $\delta\nu_{02}$ and
$\delta\nu_{13}$. The mean observed values (see Table 1) and
 their corresponding errors are represented by the large box. The grey
 box corresponds to observations without IRIS values which have been obtained
at larger solar activity. It is clear that
for our \SM\ these quantities
are too large compared to the observed values while models with microscopic diffusion
are in better agreement with the observations.
However, the models  D5 (with \mMLR) and D8 (with a
larger age) have a too small $\delta\nu_{02}$.  This quantity 
is larger than the observed one for
models  D11 and D12 where opacity is computed with a heavy element
abundance $Z_k$=\MI.
Both penetrative convection and overshooting of the core in the early 
stage of solar evolution do not modify these parameters.
Comparing the positions of models S2, D2 and  D1 respectively to those of models S3, D3, D10, we 
note that the use of OPAL \EOS\ instead of CEFF \EOS\ slightly increases
the values of both  $\delta\nu_{02}$ and $\delta\nu_{13}$. On the other hand, 
these latter models which have the best physics are  in a better agreement
with the seismic sound speed as will be seen later on. A larger age
or a smaller value of $Z/X $ could bring them in the observation box.

Globally the characteristic period
$P_0$ for g modes are lower by about 1\,minute for the \SMD\ compared to \SM.

As exhibited in Fig.~2 and in \fig{fig:ba2} to \ref{fig:ba4},
the microscopic diffusion increases the
agreement between solar models and the \SRM.
The dependence of the
normalized frequency differences
between the GONG frequencies and the theoretical frequencies
on the degree $\ell$ is very weak, leading to a small dispersion of the curves 
in Fig.~2 (right).
Globally the models D
are close to the \SRM\ within $\pm 0.2\%$ 
for the sound speed and within $\pm 1\%$ for the density.
In agreement with previous works (Gough \etal{get},
 Christensen-Dalsgaard \cite{jcd})
the sound speed is lower than in \SRM\ below the convection zone.
On the contrary, all our \SMD, but D3 and D12, present a systematic minimum
 on sound
speed differences around the radius $R\sim0.5R_\odot$. Moreover, 
below $R\sim0.4R_\odot$ all our models except models  D11, D12 have
 a sound velocity smaller than the \SRM.
The  negative  minimum around  $R\sim0.2R_\odot$ which is present in most
comparisons with inverse sound speeds, only appears
as a small depression in our model  D12
computed with OPAL equation of state and diffusion of Z with $Z_k = {\cal Z}$.

Another consequence of the microscopic diffusion, as shown in \fig{fig:hr},
is that the evolutionary
paths in the HR diagram of \SMD\ are shifted
towards higher effective temperatures compared to the paths of \SM.
\begin{figure*}
\epsfxsize =19.cm
\epsfbox[25 18 587 274]{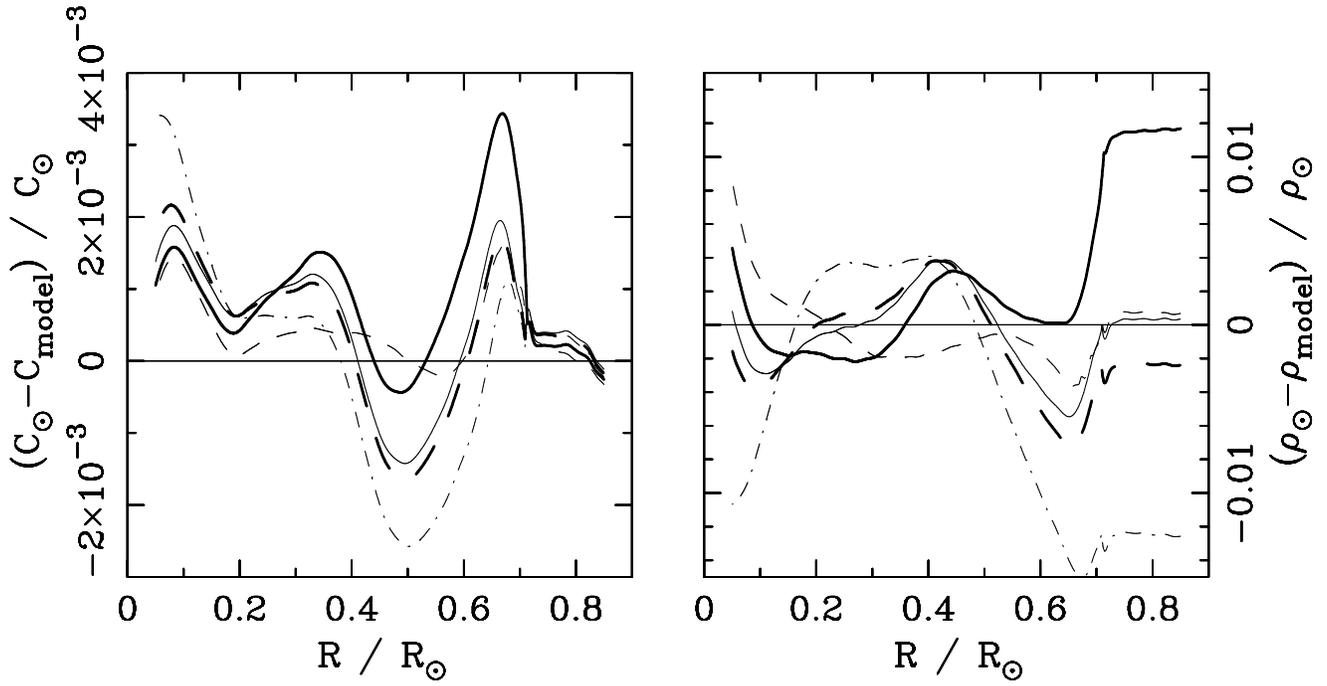}
\caption[ ]{Same as \fig{fig:ba1} for the
\SMD\ D1 \{CEFF, $Z_{\rm p}$\} ({\sl thin full}),
D2 \{CEFF, ${\cal X}_{\rm Z}$\} ({\sl thick full}),
D3 \{OPAL, ${\cal X}_{\rm Z}$\} ({\sl thin dashed}),
D4 \{CEFF, sMLR, $Z_{\rm p}$\} ({\sl thick dashed}),
D5 \{CEFF, mMLR, $Z_{\rm p}$\} ({\sl thin dash-dot-dash}).
}\label{fig:ba2} 
\end{figure*}
\begin{figure*}
\epsfxsize =19.cm
\epsfbox[25 18 587 274]{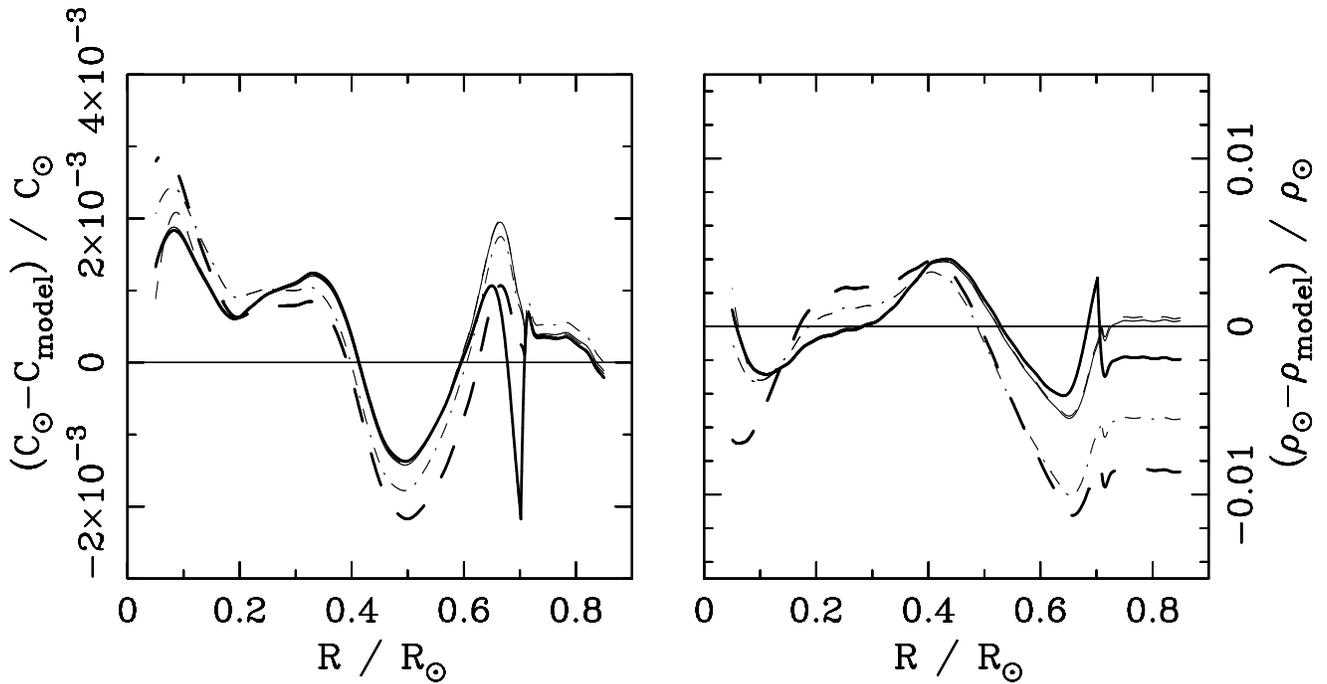}
\caption[ ]{Same as \fig{fig:ba1} for the
\SMD\ D1 \{CEFF, $Z_{\rm p}$\} ({\sl thin full}),
D6 \{CEFF, Un-shoot,  $Z_{\rm p}$\} ({\sl thick full}),
D7 \{CEFF, Ov-shoot, $Z_{\rm p}$\} ({\sl thin dashed}),
D8 \{4.65 Gy, CEFF, $Z_{\rm p}$\} ({\sl thick dashed}),
D9 \{4.65 Gy, CEFF, $Z_{\rm p}$, 0.026\} ({\sl thin dash-dot-dash}).
}\label{fig:ba3}
\end{figure*}
\begin{figure*}
\epsfxsize =19.cm
\epsfbox[25 18 587 274]{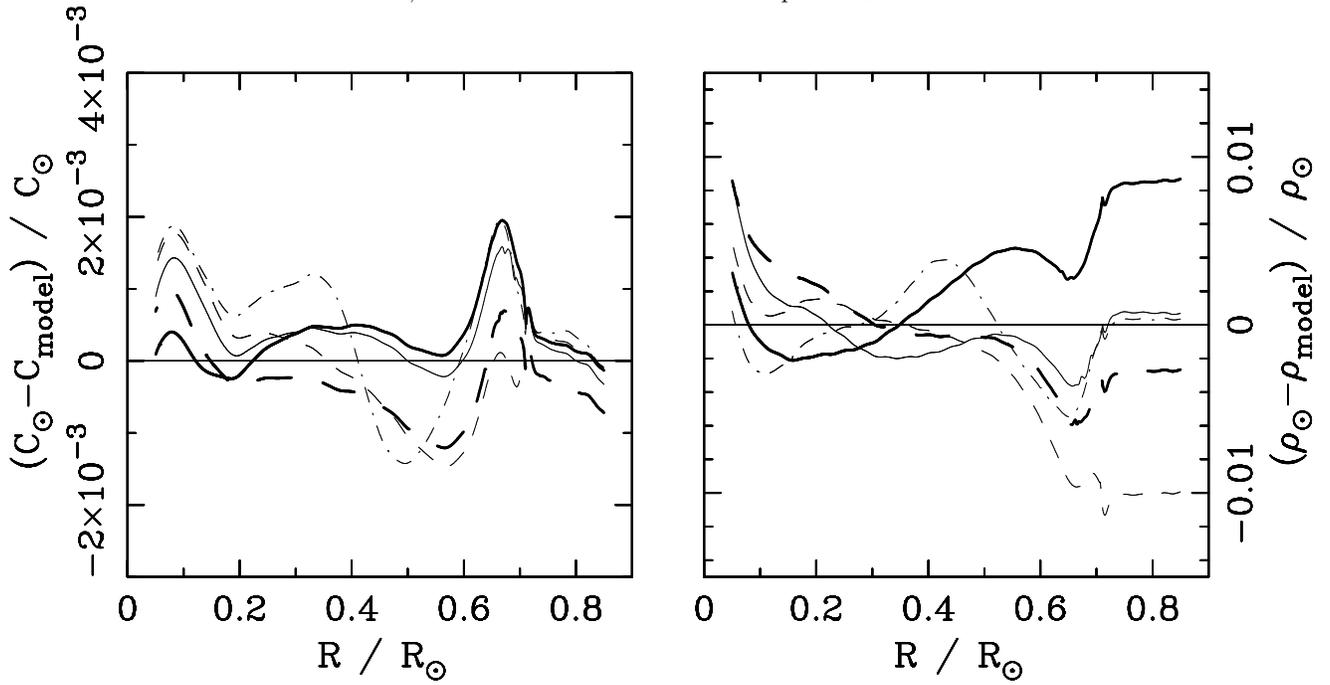}
\caption[ ]{Same as \fig{fig:ba1} for
D3 \{OPAL, ${\cal X}_{\rm Z}$\} ({\sl thin full}), 
D12 \{OPAL, ${\cal Z}$\} ({\sl thick full}),
D10 \{OPAL, $Z_{\rm p}$\} ({\sl thin dashed}),
D11 \{OPAL, sMLR, ${\cal Z}$\} ({\sl thick dashed})
and D1 \{CEFF, $Z_{\rm p}$\} ({\sl thin dash-dot-dash}); considering all 
constraints, D11 is our {\em preferred model}, but D12 without the
lithium depletion constraint.
}\label{fig:ba4}
\end{figure*}
\begin{figure}
\epsfxsize =9.cm
\epsfbox[25 18 587 774]{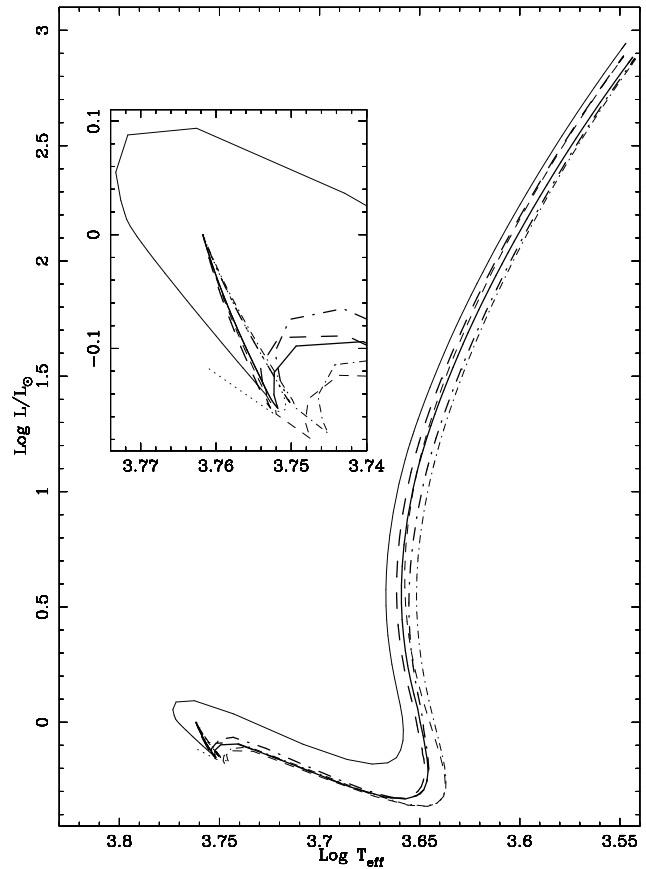}
\caption[ ]{HR diagram for models
D11 \{OPAL, sMLR, ${\cal Z}$\} ({\sl thin full}),
D12 \{OPAL, ${\cal Z}$\} ({\sl thick full}),
D1  \{CEFF, $Z_{\rm p}$\} ({\sl thin dashed}),
D10 \{OPAL, $Z_{\rm p}$\} ({\sl thick dashed}),
S2  \{CEFF, OPAL95, PMS\} ({\sl thin dash-dot-dash}),
S3  \{OPAL, OPAL95, PMS\} ({\sl thick dash-dot-dash}) and
S4  \{OPAL, OPAL95, ZAMS\} ({\sl thin dotted});
an enlargement of the main sequence is plotted in the panel.
At \ZAMS\ the models have about the same luminosities, the effective
temperatures for \SMD\ exceed by $\sim 2\%$ those of \SM.
On the main sequence the evolutionary paths of S3 and S4 are superimposed. 
}\label{fig:hr}
\end{figure}

\subsubsection{Sensitivity to \EOS\  and $Z_\kappa$}
By inspection of Figs 4, 5, 6, it is seen that the relative sound speed 
differences with \SRM, have large variations of amplitude for models
with CEFF \EOS\ with a pronounced minimum around $r=0.5R_{\odot}$. The behavior is much 
smoother for models with OPAL \EOS\ with a smaller minimum around 
 $r=0.6R_{\odot}$. This could be partly due to the way partial recombination 
in the solar radiative interior is avoided in CEFF \EOS.

The model D10 (see \fig{fig:ba4})
obtained with the \EOS\ OPAL, is located closer to the \SRM\ than the 
model D1 obtained with the \EOS\ CEFF, but 
for the density profiles, D1 is significantly closer than D10 to the \SRM. The
differences in normalized frequencies are $\sim 1\mu$Hz between the two 
models. For the models D2 and D3 , 
the amount of heavy element $Z_\kappa$ used
for the computation of \OD\ in models 
is taken as the abundance of the mean fictitious
chemical ${\cal X}_{\rm Z}$. D2 is computed with the \EOS\ CEFF and D3 with
OPAL. As seen \fig{fig:ba2}, for the sound speed D3 is closer
than D2 of the \SRM, and also for density except for $R\la 0.1R_\odot$. 
In agreement with Basu et al. (\cite{bcd}), however we find that
the closest models to the \SRM\ are obtained with the
\OD\ and \EOS\ OPAL.

Comparisons of models D3 ($Z_\kappa={\cal X}_{\rm Z}$), D10
($Z_\kappa=Z_{\rm p}=Cte$)
and D12 ($Z_\kappa=$\ \MI) allow a comparison between the three approximations
used for $Z_\kappa$. As far as the sound speed is concerned,
the models computed with an opacity value taking into account of 
the diffusion of heavy element are closer
to the \SRM\ than models with $Z_\kappa=Cte$; the models D3 and D12
are very similar, but for $R\la0.2R_\odot$, D12 is closer to the \SRM.
For the density, and $R\ga0.1R_\odot$
the model D3 is closer to the \SRM\ than D10 and D12.
Beneath the convection zone the sound velocity in model D12
is smaller than in D10 with a difference of the same
amount, but of opposite sign, of those
exhibited on Fig. 2(a) in Proffitt (\cite{pr}),
however, the physical inputs of that work differ from ours.

As a matter of conclusion, with the physical data used here, our study 
shows that the model computed with $Z_\kappa=$\ \MI\ has a 
sound speed profile closer to the \SRM\ than the model computed with
$Z_\kappa={\cal X}_{\rm Z}$, and that is reversed for the density profile.
The model D10 computed with $Z_\kappa=Cte$, though farther from the \SRM\ than
D3 and D12, remains within $\pm0.2\%$ for the sound speed and $\pm 1\%$
for the density.
Therefore, using the \SRM, we are not able us to make a real
discrimination between the three approaches employed to compute $Z_\kappa$,
nevertheless our "preferred" choice is $Z_\kappa=\ $\MI.

\subsubsection{Sensitivity to age and $(Z/X)_\odot$}
According to \tab{tab:1} the present day solar
age is known within 0.4\,Gyr, the model D8 has been evolved 0.1\,Gyr more than
the model D1 using the same physics; in the same way, model D9
has been calibrated for a larger value of $(Z/X)_\odot=0.0260$ instead of
$(Z/X)_\odot=0.0245$ for model D1. \Fig{fig:ba3} shows that
the sound speed in D9
is systematically smaller than in D8 by $\sim 0.05\%$ \ie,
about our limit of accuracy, except around the center,
while the density in D9
is greater in the convection zone and in the core.
The spread in normalized frequencies is $\sim 0.4\mu$Hz between the two models.

\subsubsection{$^7{\rm Li}$ depletion due to undershoot and mass loss}
All \SMD, except D4 and D5, reveal that the observed $^7$Li depletion does not only
result from microscopic diffusion, even if, as for model D6,
an undershooting factor equal to
the upper limit of Basu \& Antia (\cite{bba}),
namely $\zeta_{\rm un}=0.1 H_{\rm p}$,
is allowed for.
The discontinuity of the temperature
gradient due to undershooting is at the origin of spurious
bumps, though within the global accuracy, namely $\pm 0.2\%$ for
the sound speed and $\pm 0.5\%$ for density, as exhibited in \fig{fig:ba3}.
But Ahrens
\Etal{ah} have shown from models without microscopic diffusion that
an undershooting factor $\zeta_{\rm un}\gse 0.3$
is needed to fit the present day
observed lithium depletion; such a large amount is in
conflict with the upper limit for $\zeta_{\rm un}$ derived from helioseismic 
observations
(Basu \& Antia \cite{bba}; Monteiro \etal{mct}; Provost \etal{pmb}).
Our results for D6 (\tab{tab:21}) also show that the most important amount of
lithium  depletion occurs during the \PMS.
This is in conflict
with the observations
of lithium in Pleiades and Hyades (see \sec{sec:li}). The lithium being
mostly destroyed in the hot core at the end of the \PMS\ (Morel \etal{mli}) when
the young Sun is still fully convective, a larger amount of undershooting will
enhance the duration of
this fully mixed stage and increase the \PMS\ depletion.
Therefore from the properties of our models it results that
the $^7$Li depletion observed at present
day in the solar photosphere does not only result from overshooting
and microscopic diffusion. 

In contrast in the models where the mass loss is included
(models D4 and D5 of \tab{tab:21}),
the lithium depletion occurs mainly during the \MS.
The comparisons of theoretical
sound speed and the density profiles
with the \SRM\ (\fig{fig:ba3}) show that the model
D4 calculated with \sMLR\ of \eq{eq:smlr}, is closer to the
\SRM\ than the model D5 computed with \mMLR\ of \eq{eq:mmlr}.
Indeed, stronger mass loss affects the structure of the present
day model less because the mass loss phase is over faster. 
A model with \sMLR\ becomes sensitive to the mass loss at the end of the \PMS.
Therefore, the
evolutionary path in the HR diagram (\fig{fig:hr}) of the model D4 reaches the 
\ZAMS\ with
decreasing effective temperature and luminosity.
Indeed both the strong and the mild mass loss rates look as rather {\sl ad-hoc}
assumptions
so we have not attempted to adjust the free parameters of these laws
in order to fit accurately the present day observed lithium depletion.
Nevertheless it appears that a \sMLR\ is a
promising process to explain the observed lithium depletion in the Sun
as already claimed by Guzik \& Cox (\cite{gc})
despite the large expected neutrino capture rates and the marginal
agreement with 
the observed frequency difference $\delta\nu_{02}$.

\subsubsection{Overshoot of the convective core of the young Sun}
The model D7 is computed with the physics of model D1 but with an overshooting
of the convective core which appears during the early stage of solar evolution
by a factor $\zeta_{\rm ov}=0.2$. The global properties of these models 
(\tab{tab:21}) and the
comparisons with the \SRM\ reveal
only slight insignificant differences at center between D7 and D1
(\fig{fig:ba2} and \fig{fig:ba3}) except a significant increase
of the characteristic period $P_0$ of the gravity modes.

\begin{table*}
\caption[]{
Global characteristics of \SMD.
$M_{\rm p}$ is the protosolar mass, $\zeta_{\rm ov}$ and $\zeta_{\rm un}$ 
respectively are the overshooting and undershooting factors;
the other captions are given \tab{tab:11}.
}\label{tab:21}
\begin{tabular}{lllllllllllllllll} \\  \hline \\

                &  D1  &  D2  & D3   & D4   & D5   & D6   & D7   & D8   & D9   &  D10 &  D11 & D12  &\\ \\ \hline \\
$M_{\rm p}$     &  1.  &  1.  &  1.  &  1.1 & 1.095&  1.  &  1.  & 1.   &  1.  &  1.  & 1.11 &  1.  &\\
$l/H_{\rm p}$   & 2.00 & 1.95 & 1.97 & 2.00 & 2.01 & 2.00 & 2.00 & 2.02 & 2.02 & 1.97 & 1.93 & 1.92 &\\
$\zeta_{\rm ov}$&      &      &      &      &      &      & 0.2  &      &      &      &      &      &\\
$\zeta_{\rm un}$&      &      &      &      &      &  0.1 &      &      &      &      &      &      &\\
$Y_{\rm p}$     &0.276 &0.276 &0.276 &0.275 &0.273 &0.275 &0.276 &0.275 &0.279 &0.273 &0.273 &0.274 &\\
$Z_{\rm p}$     &0.0191&0.0191&0.0191&0.0191&0.0189&0.0190&0.0191&0.0191&0.0201&0.0191&0.0194&0.0196&\\
$(Z/X)_{\rm p}$ &0.0271&0.0271&0.0271&0.0270&0.0267&0.0269&0.0271&0.0271&0.0286&0.0270&0.0276&0.0277&\\
$X_{\rm nCNO}$  &0.0053&0.0053&0.0052&0.0052&0.0050&0.0051&0.0052&0.0053&0.0063&0.0052&0.0056&0.0058&\\ \\
Li$_{\rm ZAMS}$ & 3.00 & 3.01 & 2.95 & 3.13 & 3.13 & 2.74 & 3.00 & 2.99 & 2.99 & 2.88 & 3.13 & 2.95 &\\ \\
$Y_\odot$       &0.247 &0.246 &0.244 &0.247 &0.247 &0.248 &0.247 &0.246 &0.250 &0.244 &0.245 &0.245 &\\
$Z_\odot$       &0.0179&0.0179&0.0180&0.0179&0.0179&0.0179&0.0179&0.0180&0.0189&0.0180&0.0180&0.0180&\\
Li$_\odot$      & 2.92 & 2.94 & 2.88 & 2.17 & 1.96 & 2.66 & 2.92 & 2.92 & 2.92 & 2.81 & 1.92 & 2.88 &\\ \\
$R_{\rm ZC}$    &0.710 &0.713 &0.711 & 0.713& 0.709&0.702 &0.710 &0.708 &0.709 &0.707 &0.710 &0.711 &\\ \\
$T_{\rm c}$     &1.557 &1.559 &1.557 &1.558 &1.560 &1.556 &1.557 &1.560 &1.563 &1.556 &1.562 &1.565 &\\
$\rho_{\rm c}$  &151.5 &150.8 &150.2 &152.0 &153.9 &151.5 &150.8 &153.1 &156.3 &150.9 &150.6 &151.2 &\\
$Y_{\rm c}$     &0.637 &0.637 &0.635 &0.639 &0.644 &0.637 &0.634 &0.642 &0.641 &0.635 &0.639 &0.638 &\\
$Z_{\rm c}$     &0.0200&0.0201&0.0202&0.0200&0.0199&0.0199&0.0201&0.0201&0.0210&0.0201&0.0207&0.0208&\\ \\
$\Phi_{\rm Ga}$ & 128  &  128 & 128  & 136  & 140  & 128  & 128  & 129  & 130  & 127  & 144  & 130  &\\
$\Phi_{\rm Cl}$ &7.72  & 7.81 & 7.68 & 8.25 & 8.69 & 7.68 & 7.74 & 7.93 & 8.20 & 7.58 & 8.93 & 8.27 &\\
$\Phi_{\rm Ka}$ & 0.61 & 0.62 & 0.61 & 0.66 & 0.69 & 0.61 & 0.61 & 0.63 & 0.66 & 0.60 & 0.72 & 0.66 &\\ \\
$\delta\nu_{02}$& 9.00 & 9.03 & 9.10 & 8.96 & 8.85 & 9.00 & 9.00 & 8.90 & 8.96 & 9.04 & 9.13 & 9.09 &\\
$\delta\nu_{13}$&15.96 & 16.01&16.04 &15.92 &15.77 &15.97 &15.99 &15.83 &15.92 &15.98 & 15.98& 16.04&\\ \\ 
$P_{0}$         &35.73 & 35.82&35.87 & 35.63&35.27 &35.76 &35.97 &35.43 &35.65 &35.78 & 35.76& 35.75&\\
\\ \hline \\
\end{tabular}
\end{table*}

\section{Conclusions}\label{sec:dis}
We have computed solar models with our stellar evolution code CESAM and using
 updated physics.
Our models are attempts to extend the study of the sensitivity of \SMD\ to \PMS, lithium depletion, 
\ML, microscopic diffusion of heavy
species,
overshooting and undershooting. Effects of rotation and of turbulent diffusion
are ignored here.
We have compared the sound speed and the density profiles of
the \SRM\ of Basu \Etal{bcd} to 
calibrated solar model computed with various opacity data,
equation of state, microscopic diffusion, mass loss rate, undershooting
and overshooting amounts. The \SMD\ agree
with the seismic model within $\pm 0.2\%$ for the sound speed and
$\pm 1\%$ for density, while for the \SM\ the
agreement is hardly better than $\pm1\%$ and $\pm10\%$ respectively
for the sound speed and density. For the \SMD\ the
depth of the convection zone and the amount of helium at surface agree
fairly well with their values inferred by helioseismology.
A significant increase of the quality of solar models results from 
the recent improvements of \OD\ and \EOS,
any amelioration of solar models
will necessitate to take fully into account the changes of
chemicals in \OD\ and \EOS.

Our models reveal that
the lithium depletion observed at solar surface is certainly not due to
undershooting at the bottom of the convection zone and can be explained by
a strong mass loss occurring during the first 200\,Myr; however, in the solar
models described in this paper
the turbulent diffusion induced either by the rotation 
or by the internal waves
 has not been taken into account.
 
Considering all the observational and helioseismologic constraints
our {\em preferred model} is D11 computed with \sMLR, OPAL \OD\ and \EOS,
microscopic
diffusion of hydrogen, helium and, as trace elements, all the CNO
species plus a fictitious mean no-CNO chemical which models the
microscopic diffusion of the heaviest elements of the mixture; but
if one excludes the constraint on the lithium depletion, our
{\em preferred model}
is D12 computed with the same physics but without mass loss.
However, in the present state of art, the precision
of solar models needs to be still improved by
one magnitude in order to reach the accuracy of the solar data
inferred by helioseismology.

\begin{acknowledgements}
We are grateful to G. Alecian, M. Gabriel, S. Brun and F. Th\'evenin
for helpful
discussions, to G. Houdek for providing the opacity interpolation package
and to N. Audard for help to include this package in CESAM code.
We thank S. Basu for her seismic model and
the GONG project for providing p-modes frequencies.
We want to express our thanks to Dr. J. Guzik who referred this paper 
whose comments and remarks and english corrections
greatly helped to improve the presentation of this paper.
This work was partly supported by the GDR G131 "Structure Interne" of CNRS
(France).
\end{acknowledgements}


\appendix
\section{Calculation of abundances}\label{ap:1}
The \OD\ and \EOS\ are tabulated for mixtures with various amounts
of hydrogen $X_{\rm H}$, and heavy element content $Z$;
the ratios between the species of which $Z$ is made of are fixed;
along the solar evolution these ratios are modified by 
thermonuclear reactions and microscopic diffusion.
Therefore, using the available data for \OD\ and \EOS, a difficulty is the 
estimate of $Z$ as consistently as possible.
At time $t$, $X_i$, the amount per unit of mass of the species labelled with
$i=1,\ldots, N_c$ ($N_c$ is the total number of chemicals) is written:
\[
X_i=\frac{n_i M_i m_u}{\sum_{j=1}^{N_c} n_j M_j m_u}=\frac{n_i M_i m_u}\rho
\]
and
\begin{equation}\label{eq:z1}
1\equiv\sum_{j=1}^{N_c}X_i=X_\ncno+\sum_{j=1}^N X_i
\end{equation}
here, $n_i$ is the number density, $M_i$ is the atomic mass of $X_i$,
$M_i\in\bbbr$ differs from the {\em integer}
atomic number $A_i$, (Clayton \cite{c}),  $m_u$
is the atomic mass unit, $N\leq N_c$ is the number of species
entering into the nuclear network, $X_\ncno$ is the amount, per mass unit, of 
a fictitious mean heavy element of mean atomic mass
$M_\ncno$ which does not belong to the nuclear network and $\rho$ is the 
density.
Therefore $Z$ is  writen:
\[
Z\equiv 1-X_{\rm H} - X_{\rm He}=X_\ncno+
\sum_{j\not={\rm H,He}} X_j;
\]
for sake of clarity the labels "H" and "He" are used in place
of integer indexes;  $X_{\rm H}$ and $X_{\rm He}$ also
include, respectively, the isotopes of hydrogen and helium, (see \App{ap:2}).
Following Fowler \Etal{fcz}, we use the number of mole\,g$^{-1}$,
$Y_i\equiv X_i/M_i$; then 
the amount, per unit of mass, of the species indexed with $i$ is written:
\[
X_i=\frac{Y_i M_i}{Y_\ncno M_\ncno +\sum_{j=1}^N Y_j M_j},
\]
namely the abundances of hydrogen,
and heavy element which are entries of \OD\ or \EOS, are respectively written:
\begin{eqnarray} \label{eq:z}
X_{\rm H}&=&\frac{Y_{\rm H} M_{\rm H}}
{Y_\ncno M_\ncno +\sum_{j=1}^N Y_j M_j}, \nonumber \\
Z&=&\frac{Y_\ncno M_\ncno+\sum_{j\not={\rm H,He}} Y_j M_j}
{Y_\ncno M_\ncno +\sum_{j=1}^N Y_j M_j}
\end{eqnarray}
therefore \eq{eq:z1} is fullfilled, despite the fact that the quantity
which is conserved is the nucleon number expressed as:
\[
0\equiv\sum_{i=1}^NA_i\ddx{X_i}t=\sum_{i=1}^NA_iM_i\ddx{Y_i}t,
\]
The similar approach of Richard \Etal{rvcd} uses the normalization equation:
\[
1\equiv X_{\rm H}+\sum_{i\not= {\rm H}} X_i+Z.
\]
\begin{figure}
\epsfxsize = 9cm
\epsfbox[25 18 530 700]{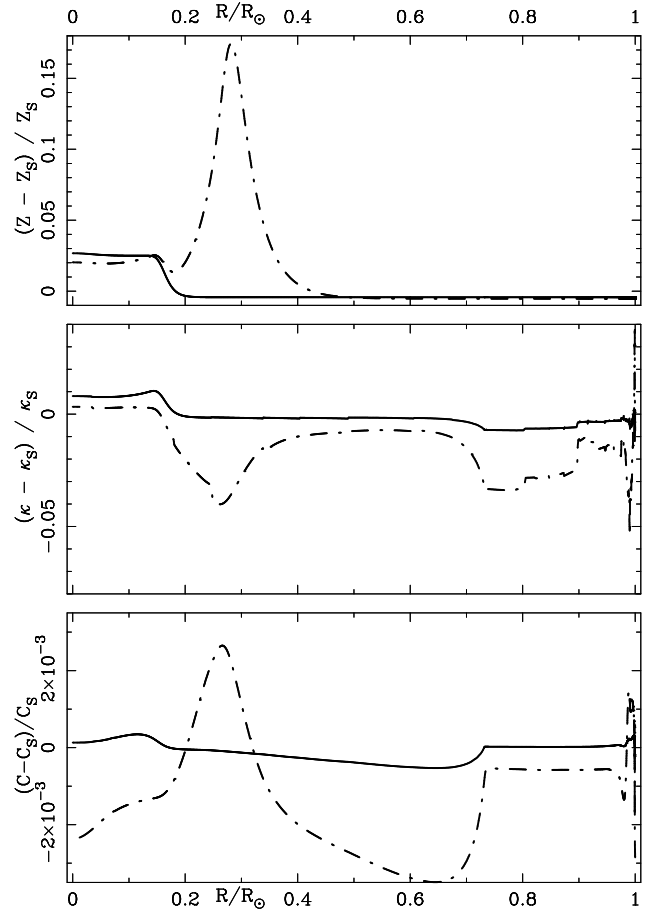}
\caption[ ]{
Relative differences for heavy element contents $Z$ (top), $\kappa$
opacities (middle) and $C$ sound velocities 
(bottom) along the normalized radius, 
between the model S1 and \SM\ in which $^3$He either, {\em is}, Sy ({\sl full})
or, {\em is not}, Sz ({\sl dash-dot-dash}), included into the helium
content for \OD\ calculations.
}\label{fig:1}
\end{figure}

\section{$^3$He abundance, \EOS\ and \OD}\label{ap:2}
\begin{figure}
\epsfxsize = 8cm
\epsfbox{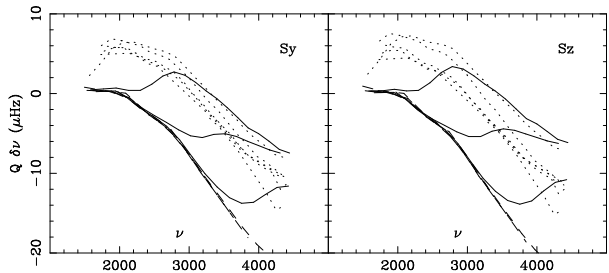}
\caption[ ]{O-C frequencies between sets of modes of constant $\ell$ computed 
between GONG observations and \SM\ computed respectively with,
$Z=1-^1$H$-^4$He (Sz) and $Z\equiv 1-^1$H$-^3$He$-^4$He (Sy).
}
\label{fig:Z}
\end{figure}

In \MS\ solar models, a local maximum of $^3$He, occurs around
$R\sim 0.3R_\odot$,
it amounts to 10\% of the local helium content, there $X_{^3{\rm He}}\sim Z$;
this effect needs to be
taken into account in the calculation of $Z_\kappa$; \fig{fig:1} exhibits
the fractional differences on $Z_\kappa$, opacity and sound velocity
between the model S1 used as a reference -- for S1,
$Z_\kappa\equiv Cte$ (see \tab{tab:31}) -- and
\SM\ (not detailed here for sake of briefness) calculated {\em with} and
{\em without} $^3$He included into the helium content \ie,
$Z_\kappa\equiv 1-^1$H$-^3$He$-^4$He and $Z_\kappa\equiv1-^1$H$-^4$He. 
From
\fig{fig:1} the fractional differences on sound speeds are of the order
of $\pm 3\times10^{-3}$ \ie, of the order of accuracy achieved with \SMD.
Likewise \fig{fig:Z} shows that the spread of O-C frequencies between sets of
modes of constant $\ell$ computed between GONG observations 
decreases by taking (left) $Z\equiv 
1-^1$H$-^3$He$-^4$He instead of (right) $Z\equiv 1-^1$H$-^4$He.
The difference of the sound speed below the convection zone
and around $R\sim 0.3R_\odot$ seen in \fig{fig:1} induces scaled frequency
difference between the two models of the order of $1\,\mu$Hz, as seen in Fig.
10.
\begin{figure}
\epsfxsize = 8cm
\epsfbox{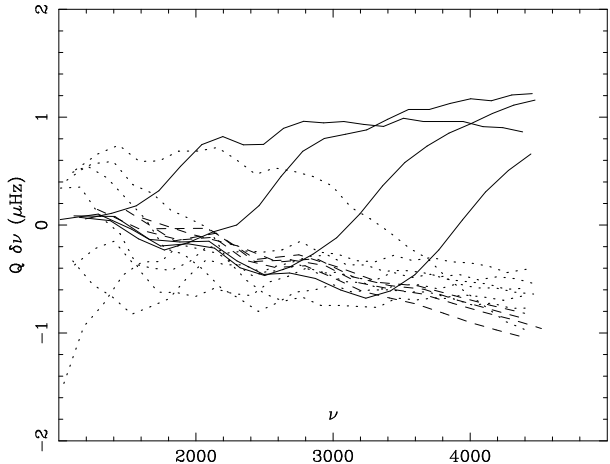}
\caption[ ]{Scaled frequencies differences between the models Sy and Sz
of \fig{fig:Z} for modes of low degree ($\ell$= 0, 1, 2, 3, 4, 5, 10;
{\sl dotted}), intermediate degree ($\ell$= 20, 30, 40, 50; {\sl full})
and high degree ($\ell$= 70 100 120 140; {\sl dashed}).
}
\label{fig:ja}
\end{figure}

Everywhere, but around $\simeq 0.3R_\odot$, the small abundance of $^3$He
has no effect on \OD\ and
\EOS; around  $R\simeq 0.3R_\odot$ \OD\ and \EOS\ are mainly sensitive
to free-free 
transitions, there $^3$He and $^4$He are equivalent providers of electrons;
therefore it is relevant to include $^3$He
in the helium content for the calculation of \OD.
Similar phenomenon prevails for all components of $Z$
which present noticeable 
changes of abundances due to nuclear reactions and diffusion \eg, for $^{12}$C 
and $^{14}$N if $R\lse 0.2R_\odot$.
The sensitivity of the sound velocity to such effect being of same order of 
the accuracy reached by \SMD, we 
emphasize that any improvement of that accuracy will necessitate 
\OD\ -- even \EOS\ -- taking into account the changes of mixture.

\end{document}